\documentclass[%
 reprint,
 amsmath,amssymb,
 aps,
]{revtex4-2}

\usepackage{graphicx}
\usepackage{dcolumn}
\usepackage{bm}
\usepackage{gensymb}
\usepackage{textcomp}

\begin{document}

\preprint{APS/123-QED}

\title{Universal Scaling of Freezing Morphodynamics in Polymer Solution Droplets}

\author{Nicolas G. Ulrich}
\author{Pravin P. Aravindhan}
\author{Olivia Berger}
\author{Bryan S. Beckingham}
\author{Jean-François Louf}%
 \email{jlouf@auburn.edu}
\affiliation{%
 Chemical Engineering, Auburn University \\
 202 Ross Hall, Auburn AL 36849 
}%


\date{\today}

\begin{abstract}
Freezing of complex fluids is central to a wide range of natural and technological processes, where the interplay between heat transport, solute redistribution, and interfacial deformation gives rise to complex morphologies. Unlike simple liquids, polymer solutions exhibit strongly coupled transport and rheological properties that evolve dynamically during solidification, making their freezing behavior difficult to predict. Here, we examine the freezing of polymer solution droplets spanning dilute to entangled regimes. We find that droplet morphology and freezing dynamics in viscous solutions are governed by a single dimensionless parameter, the Capillary--Lewis number, which captures the competition between viscous stresses, capillarity, and solute transport. Circularity, radial deformation, and freezing time collapse onto a master curve spanning nine orders of magnitude, revealing a transition near unity corresponding to the point at which solute diffusion can no longer relax concentration gradients ahead of the freezing interface. This collapse holds across distinct polymer chemistries within the viscous fluid regime, while deviations emerge when the material exhibits elastic-dominated response ($G' > G''$), indicating the breakdown of purely transport--capillary control. These results establish a minimal transport--mechanics framework linking solute redistribution to interfacial deformation during freezing polymer solutions.
\end{abstract}

\maketitle


Freezing droplets often produce complex morphologies that emerge from the coupling of heat transport, phase change, and fluid motion during solidification \cite{wildeman2017fast, kuznetsov2019marangoni,khawaja2025exploring,karlsson2019experimental}. Even in simple liquids such as pure water, the propagation of the freezing front redistributes mass and heat, generating characteristic droplet shapes as the solid phase advances \cite{snoeijer2012pointy, jambon2018singular, zhang2019shape,seguy2023role,marin2014universality}. When solutes are present, this process becomes significantly more complex because the advancing interface rejects molecules into the remaining liquid\cite{albouy2017freezing}, creating concentration gradients that modify both the transport properties and the rheology of the fluid \cite{chu2024interfacial,fan2018phase,ulrich2026interfacial}.

Polymer solutions provide a particularly rich system for exploring these coupled effects. Polymer concentration and molecular weight strongly influence viscosity and molecular diffusivity \cite{dobrynin2023viscosity,luo2009scaling}, while rejected polymer chains accumulate near the freezing front as solidification proceeds \cite{you2019interactions,wegst2010biomaterials}. This local enrichment increases the viscosity of the surrounding liquid and alters the fluid response to interfacial stresses \cite{zhang2021planar,wang2016interface}. As a result, the macroscopic morphology of freezing polymer droplets reflects a dynamic interplay between solute redistribution, viscous stresses, and capillary forces \cite{wu2012preparation}.

Previous studies have shown that freezing polymer-containing systems can produce complex structures and shape transformations arising from solute rejection and transport-limited solidification \cite{yin2023hierarchical,wang2021three,wang2016interface}. Experiments with polymer solutions such as poly(vinyl alcohol) demonstrate that increasing polymer concentration modifies freezing dynamics and droplet morphology by altering transport and rheological properties of the liquid phase \cite{kharal2023unidirectional}. Related effects occur in other solute-driven freezing systems, including antifreeze protein solutions, where molecular interactions with the advancing ice interface influence front propagation and morphology \cite{naullage2018controls, naullage2020slow,kim2020novel,sander2004kinetic,chasnitsky2019ice}. Despite these advances, a general framework linking molecular transport in polymer solutions to the macroscopic morphologies of freezing droplets remains lacking.

Here, we investigate the freezing of polymer solution droplets over a wide range of molecular weights and concentrations spanning dilute to entangled regimes. By independently measuring viscosity, surface tension, thermal diffusivity, and freezing dynamics, we show that freezing morphodynamics of viscous polymer solutions follow a universal scaling governed by a single dimensionless parameter, the Capillary–Lewis number, which captures the competition between viscous deformation, capillary resistance, and diffusive solute transport during front propagation. This scaling reveals a sharp transition between diffusion-dominated and viscous–capillary regimes and is robust across distinct polymer chemistries within the viscous fluid regime, providing a predictive framework linking polymer-solution transport to interfacial solidification. Deviations emerge when the material exhibits elastic-dominated behavior, indicating the limits of transport–capillary control.

Droplets of Poly(ethylene glycol) diacrylate (PEGDA), synthesized via standard acrylation of PEG (see Supplemental Material for details on PEGDA synthesis; Ref.~\cite{hahn2006photolithographic}), and poly(vinyl alcohol) (PVA) solutions with controlled molecular weight and concentration were deposited on a -20$\degree$C substrate and frozen while imaging the advancing freezing front. Polymer solutions spanned dilute, semi-dilute, and entangled regimes \cite{graessley1980polymer,de1979scaling}, allowing viscosity to vary over four orders of magnitude while diffusivity varied by approximately one order of magnitude \cite{luo2009scaling,ram2025enhanced,holyst2009scaling, mintis2019effect}. Surface tension, viscosity, and thermal diffusivity were independently measured, and for PVA, rheological measurements were additionally performed to identify the transition from viscous to gel-like behavior (see Supplemental Material for experimental protocols and analysis methods). The evolution of droplet morphology and freezing front position was tracked from optical imaging.

During freezing, the advancing solidification front progressively deforms the liquid droplet. A representative time sequence is shown in Fig.~1(a). In contrast to pure water droplets, which typically develop a pointed tip as freezing proceeds \cite{snoeijer2012pointy}, polymer-containing droplets exhibit a wide range of final shapes depending on polymer concentration and molecular weight \cite{zhao2021freezing,kharal2023unidirectional}.

\begin{figure}
\includegraphics{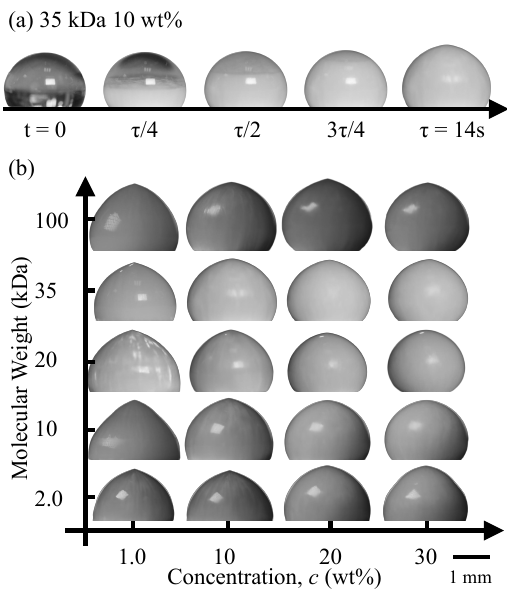}
\caption{\label{fig:morph} Morphology map of PEGDA solution droplets frozen on a flat substrate at $-20^\circ$C as a function of polymer concentration and molecular weight. Representative final droplet shapes are shown for each condition, illustrating the evolution from pointed to rounded geometries with increasing concentration and molecular weight. Representative droplets are shown without geometric normalization; variations in initial droplet size arise from viscosity-dependent detachment and do not affect the observed morphological trends.}
\end{figure}

Systematically varying polymer concentration and molecular weight reveals a clear morphological transition. The resulting frozen droplet geometries are summarized in the morphology map of Fig.~1(b). At low polymer content, the droplets develop the sharply pointed morphology characteristic of freezing water droplets. Increasing either concentration or molecular weight progressively suppresses this tip formation and modifies the overall droplet geometry, producing smoother and more rounded dome-like shapes. This evolution indicates that solute transport within the liquid and the associated viscous stresses significantly influence the fluid response during front propagation.

To identify the mechanisms underlying this transition, we next examine the dynamics of the freezing front. The evolution of the front position for droplets of 35 kDa and increasing polymer concentration is shown in Fig.~2. As polymer content increases, the propagation of the freezing front slows markedly and the total freezing time increases, indicating that transport within the liquid phase plays a key role in controlling the solidification process.

\begin{figure} [htb]
\includegraphics{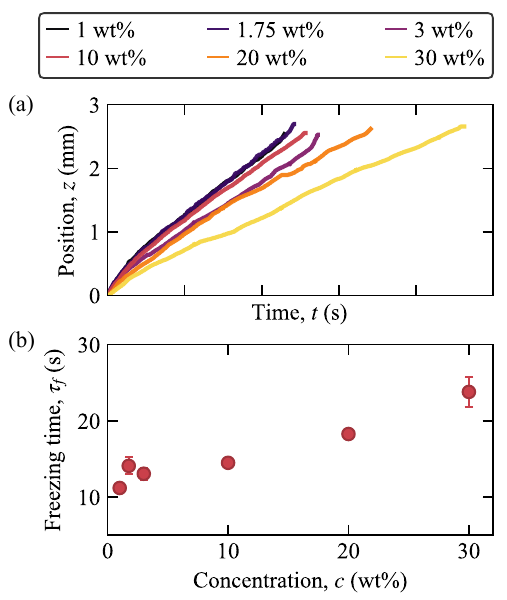}
\caption{\label{fig:dynamics} (a) Freezing front position $z$ as a function of time for 35 kDa PEGDA droplets at increasing concentrations. (b) Corresponding total freezing time $\tau$ as a function of concentration.}
\end{figure}

As solute accumulates near the interface, the resulting increase in viscosity enhances resistance to flow, limiting liquid redistribution toward the droplet apex and suppressing the formation of the pointed tip characteristic of pure liquids.

To quantify these shape changes, we extracted geometric descriptors from the final frozen droplets. Figure~3(a) shows the circularity $\chi$ as a function of polymer concentration for a fixed molecular weight (35 kDa). Circularity increases monotonically with concentration before reaching a plateau at higher concentrations, reflecting the progressive suppression of the pointed morphology and the transition toward more isotropic droplet shapes.

\begin{figure} [htb]
\includegraphics{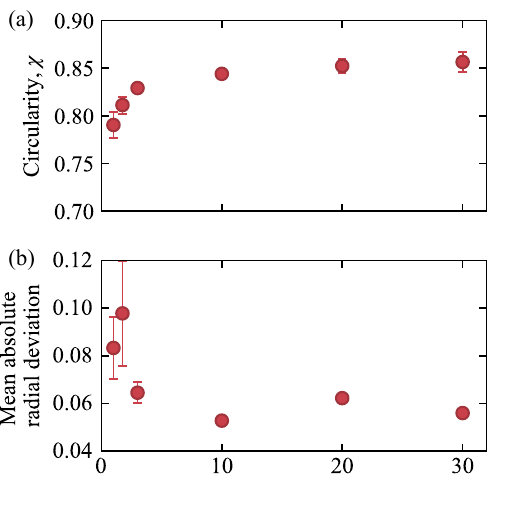}
\caption{\label{fig:conc} Morphological metrics for frozen droplets containing 35 kDa PEGDA at varying concentrations. (a) Circularity $\chi$ as a function of concentration. (b) Mean absolute radial deviation quantifying deviations from a circular shape.}
\end{figure}

A complementary metric, the mean absolute radial deviation (MARD) (Fig.~3(b)), defined as the spatially averaged absolute radial deformation, provides a scalar measure of the angular deformation field reported in Ref.~\cite{seguy2025freezing} (see Supplemental Material for definition and extraction). MARD decreases monotonically with concentration before plateauing, confirming the same transition toward reduced anisotropic deformation. The agreement between these two independent measures shows that the observed morphological evolution is robust and not dependent on the specific choice of geometric descriptor.

Together with the front-propagation measurements in Fig.~2, these observations indicate that both the freezing dynamics and the resulting morphology are governed by transport processes within the liquid phase during solidification.

\begin{figure} [ht!]
\includegraphics{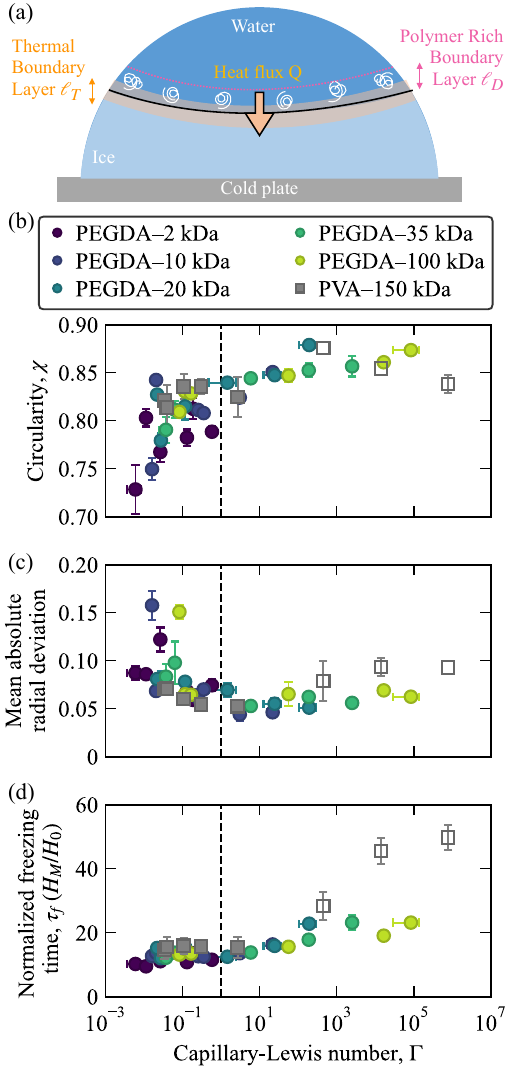}
\caption{\label{fig:gamma1} Schematic and scaling of freezing morphodynamics. (a) The advancing front (velocity $U$) generates thermal and solute boundary layers $\ell_T \sim \alpha/U$ and $\ell_D \sim D/U$, whose relative magnitude controls solute accumulation and interfacial deformation.  (b) Circularity increases with $\Gamma$ until plateauing, reflecting the transition toward more isotropic droplet shapes. (c) Mean absolute radial deviation decreases with $\Gamma$ until plateauing, indicating suppression of anisotropic deformation. (d) Size-corrected freezing time $\tau_f (H_M/H_0)$ remains approximately constant for $\Gamma \lesssim 1$ and then increases sharply, reflecting transport-limited front propagation. Filled symbols correspond to viscous solutions ($G'' > G'$), while open symbols indicate elastic-dominated conditions ($G' > G''$).}
\end{figure}

To rationalize the coupled evolution of freezing dynamics and droplet morphology, we introduce a dimensionless parameter that captures the competition between viscous deformation and transport processes during solidification. The advancing freezing interface moves with a characteristic velocity $U$, generating viscous stresses of order $\mu U/R$ in the liquid, opposed by capillary stresses $\gamma/R$, where $\mu$ is the viscosity, $\gamma$ the surface tension, and $R$ the droplet radius. Meanwhile, freezing is governed by the transport of heat and rejected polymer molecules ahead of the interface. The relative efficiency of thermal and solute transport is quantified by the Lewis number $Le=\alpha/D$, where $\alpha$ is the thermal diffusivity (measured independently) and $D$ the polymer diffusivity, obtained from literature measurements and viscosity-based scaling (see Supplemental Material for details on diffusivity estimation; diffusion coefficients from Refs.~\cite{luo2009scaling,hong2001solvent}). These coupled processes define a Capillary–Lewis number
\begin{equation}
\Gamma = Ca\times Le = \left(\frac{\mu U}{\gamma}\right) \times \left(\frac{\alpha}{D}\right),
\end{equation}
which compares viscous stresses generated during freezing to the ability of solute diffusion to relax concentration gradients relative to heat transport.

The form of this scaling can be understood by considering the coupled processes governing front propagation (Fig.~4a). The freezing velocity $U$ is set by thermal transport, scaling as $U \sim \alpha/\ell_T$, where $\ell_T$ is the thermal boundary layer thickness. In contrast, the redistribution of polymer chains occurs over a diffusive length scale $\ell_D \sim D/U$. When $\ell_D \gg \ell_T$, diffusion efficiently redistributes solute and concentration gradients remain weak, whereas when $\ell_D \lesssim \ell_T$, solute accumulates near the interface. The competition between these transport processes, quantified by $Le=\alpha/D$, acts in concert with viscous stresses opposed by capillarity, naturally leading to the Capillary--Lewis number as the governing parameter.

When the freezing dynamics and droplet morphologies are expressed in terms of the Capillary--Lewis number, the data collapse onto a universal curve spanning nine orders of magnitude. As shown in Fig.~4b--d, circularity, radial deformation, and size-corrected freezing time all vary systematically with $\Gamma$ for PEGDA solutions. Because droplet detachment from the dispensing needle introduces modest, viscosity-dependent variations in initial droplet size \cite{banitabaei2016pneumatic,rothert2003formation}, the freezing-time data in Fig.~4(d) were rescaled using the initial droplet height (see Supplemental Material for normalization procedure). This correction preserves the same $\Gamma$-dependent trend, demonstrating that the observed scaling is not an artifact of size variations.

The morphological trends observed in Fig.~1 and the slowing of the freezing front in Fig.~2 are organized by this scaling. The crossover at $\Gamma \sim 1$ emerges when the diffusive length scale $\ell_D \sim D/U$ becomes comparable to the thermal boundary layer $\ell_T \sim \alpha/U$, such that solute redistribution can no longer relax concentration gradients ahead of the interface. In this limit, solute accumulation generates viscous stresses that compete directly with capillarity, marking the transition from diffusion-dominated freezing to a viscous–capillary regime that governs droplet deformation. This transition is directly visualized by the ordered droplet morphologies in Fig.~5 and marks a transition from near-equilibrium interface evolution to a regime where transport limitations generate nonequilibrium stresses that control the macroscopic geometry. Thus, $\Gamma \sim 1$ defines a physically meaningful boundary where transport-limited solute accumulation becomes dynamically coupled to interfacial mechanics.

\begin{figure} [htb]
\includegraphics{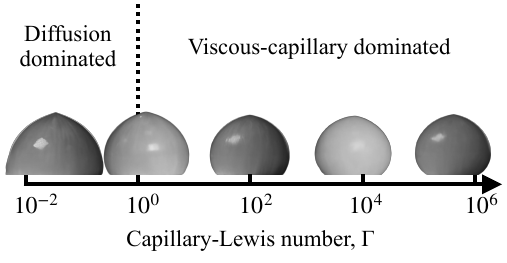}
\caption{\label{fig:heatmap} Representative frozen droplet shapes from Fig. 1 reordered by increasing $\Gamma$ ($10^{-2} \le \Gamma \le 10^{6}$). Droplets are rescaled to a common initial height for visualization. A transition near $\Gamma \sim 1$ separates diffusion-dominated freezing from a viscous–capillary regime, where solute accumulation generates stresses that suppress interfacial deformation.}
\end{figure}

To test the generality of this scaling beyond PEGDA, we performed the same analysis on PVA solutions with molecular weight 150 kDa, for which concentration-dependent cooperative diffusion coefficients are available in the literature \cite{hong2001solvent}. The PVA data collapse onto the same master curve for the morphological metrics across concentrations, including conditions where the material exhibits elastic-dominated behavior ($G' > G''$). 

In contrast, deviations emerge in the freezing time at high concentrations, where the measured values lie systematically above the $\Gamma$-dependent scaling. Independent rheological measurements indicate that this regime corresponds to $G' > G''$, marking the onset of elastic-dominated response (see Supplemental Material). This suggests that while interfacial morphology remains primarily governed by transport–capillary coupling, the freezing dynamics become increasingly influenced by elastic stresses that are not captured by the viscous scaling underlying $\Gamma$. 

In summary, we show that the freezing morphodynamics of polymer solution droplets are governed by a single dimensionless parameter combining capillary stresses and the relative transport of heat and solute. Across a wide range of polymer concentrations and molecular weights, droplet morphology collapses when expressed in terms of the Capillary--Lewis number, revealing a transition near $\Gamma \sim 1$. While freezing dynamics follow the same scaling in viscous solutions, deviations emerge in elastic-dominated regimes, indicating the limits of purely transport–capillary control. 

More broadly, this work highlights how the interplay between thermal and solute transport can be captured through minimal dimensionless groups to predict complex interfacial morphologies in multiphase systems. These results provide a predictive framework for freezing in complex fluids and may inform processes such as ice templating, cryopreservation, and polymer processing where solute transport and interfacial deformation are coupled during solidification.

We acknowledge support from the ACS PRF DNI grant $\#$67719-DNI7.

\bibliography{ref}
\nocite{schindelin2012fiji}
\end{document}